\documentclass[aps,twocolumn,showpacs,preprintnumbers,amsmath,amssymb,nofootinbib,superscriptaddress,showkeys]{revtex4}

\usepackage{epsfig}
\usepackage{graphicx}

\begin{document}

\title{Low scale saturation of Effective NN Interactions and their
  Symmetries}
%
%
\author{E. Ruiz
  Arriola}\email{earriola@ugr.es}
  \affiliation{Departamento de F\'isica At\'omica, Molecular y Nuclear,
  Universidad de Granada,
  E-18071 Granada, Spain.}
\date{\today}

\begin{abstract} 
The Skyrme force parameters can be uniquely determined by coarse
graining the NN interactions at a characteristic momentum scale.  We show how
exact $V_{\rm low k}$ potentials to second order in momenta are
accurately and universally saturated with physical NN scattering
threshold parameters at CM momentum scales of about $\Lambda=250 {\rm
  MeV}$ for the S-waves and $\Lambda=100 {\rm MeV}$ for the
P-waves. The pattern of Wigner and Serber symmetries
unveiled previously is also saturated at these scales. 
\end{abstract}

\maketitle



The derivation of effective interactions from NN dynamics has been a
major task in Nuclear Physics ever since the pioneering works of
Moshinsky~\cite{Moshinsky195819} and Skyrme~\cite{Skyrme:1959zz}. The
use of those effective potentials, referred to as Skyrme forces, in
mean field calculations can hardly be exaggerated due to the enormous
simplifications that are implied as compared to the original many-body
problem~\cite{Vautherin:1971aw,Negele:1972zp,Chabanat:1997qh,Bender:2003jk}.
Similar ideas advanced by Moszkowski and
Scott~\cite{1960AnPhy..11...65M} have become rather useful in Shell
model calculations~\cite{Dean:2004ck,Coraggio:2008in}.  The Skyrme
 (pseudo)potential is usually written in coordinate space and
contains delta functions and its derivatives~\cite{Skyrme:1959zz}. In
momentum space it corresponds to a power expansion in the CM momenta
$({\bf p'}$ and $ {\bf p})$ corresponding to the initial and final
state respectively. To second order in momenta the potential reads
\begin{eqnarray} 
&& V ({\bf
    p}',{\bf p}) 
= \int d^3 x e^{-i {\bf x}\cdot ({\bf p'}-{\bf p})}  V({\bf x} ) 
 \nonumber \\ &=&  t_0 (1 + x_0 P_\sigma ) + \frac{t_1}2(1 + x_1
  P_\sigma ) ({\bf p}'^2 + {\bf p}^2)  \nonumber \\ &+& 
 t_2 (1 + x_2
  P_\sigma ) {\bf p}' \cdot {\bf p} + 2 i W_0 {\bf S} \cdot({\bf p}'
  \wedge {\bf p}) \nonumber \\ &+& 
\frac{t_T}2 \left[ \sigma_1 \cdot {\bf p}
  \, \sigma_2 \cdot {\bf p}+ \sigma_1 \cdot {\bf p'} \, \sigma_2
  \cdot {\bf p'} - \frac13 \sigma_1 \, \cdot 
\sigma_2 ({\bf p'}^2+  {\bf p}^2)
\right]  \nonumber \\ &+& 
\frac{t_U}2 \left[ \sigma_1 \cdot {\bf p}
  \, \sigma_2 \cdot {\bf p}'+ \sigma_1 \cdot {\bf p'} \, \sigma_2
  \cdot {\bf p} - \frac23 \sigma_1 \, \cdot 
\sigma_2 {\bf p'}\cdot  {\bf p}
\right] 
\label{eq:skyrme2}
\end{eqnarray} 
where $P_\sigma = (1+ \sigma_1 \cdot \sigma_2)/2$ is the spin exchange
operator with $P_\sigma=-1$ for spin singlet $S=0$ and $P_\sigma=1$
for spin triplet $S=1$ states. In practice, these effective forces are
parameterized in terms of a few constants which encode the relevant
physical information and should be deduced directly from the
elementary and underlying NN interactions.  Unfortunately, there is a
huge variety of Skyrme forces depending on the fitting strategy
employed (see e.g. \cite{Friedrich:1986zza,Klupfel:2008af}). This lack
of uniqueness may indicate that the systematic and/or statistical
uncertainties within the various schemes are not accounted for
completely. Interestingly, the natural units for those parameters have
been outlined in Ref.~\cite{Furnstahl:1997hq,Kortelainen:2010dt}
yielding the correct order of magnitude. A microscopic
basis~\cite{Baldo:2010nx,Stoitsov:2010ha} for the Density Functional
Theory (DFT) approach has also been set up, but still uncertainties
remain.

Although the pseudo-potential in Eq.~(\ref{eq:skyrme2}) may be taken
literally in mean field calculations, due to the finite extension of
the nucleus, its interpretation in the simplest two-body problem
requires some regularization to give a precise meaning to the Dirac
delta interactions. The standard view of a pseudo-potential (in the
sense of Fermi) is that it corresponds to the potential which in the
Born approximation yields the real part of the full scattering
amplitude. This is a prescription which implements unitarity, but
necessarily fails when the scattering length is unnaturally large as
it is the case for NN interactions. On the contrary, the Wilsonian
viewpoint corresponds to a coarse graining of the NN interaction to a
certain energy scale. There are several schemes to coarse grain interactions in
Nuclear Physics.  The traditional way has been by using the oscillator
shell model, where matrix elements of NN interactions are evaluated
with oscillator constants of about $b=1.4-2 \,{\rm
  fm}$~\cite{Coraggio:2008in}. A modern way of coarse graining nuclear
interactions is represented by the $V_{\rm low k}$
method~\cite{Bogner:2001gq} (for a review see \cite{Bogner:2009bt})
where all momentum scales above $2 {\rm fm}^{-1}$ are integrated
out. The recent Euclidean Lattice Effective Field Theory (EFT)
calculations~(for a review see e.g. \cite{Lee:2008fa}), although
breaking rotational symmetry explicitly, provide a competitive scheme
where coarse grained interactions allow ab initio calculations
combining the insight of EFT and Monte-Carlo lattice experience, with
lattice spacings as large as $a=2 {\rm fm}$. These length scales match
the typical inter-particle distance of nuclear matter $d = 1
/\rho^{\frac13} \sim 2 {\rm fm}$. Actually, the three approaches
feature energy-, momentum- and configuration space coarse graining
respectively and ignore explicit dynamical effects below distances
$\sim b \sim 1/\Lambda \sim a $ which advantageously sidestep the
problems related to the hard core and confirm the modern view that
{\it ab initio} calculations are subjected to larger systematic
uncertainties than assumed hitherto. Clearly, any computational set up
implementing the coarse graining philosophy yields by itself a {\it
  unique} definition of the effective interaction. However, there is
no universal effective interaction definition.  For definiteness, we will follow here the $V_{\rm low k}$ scheme to
 determine the effective parameters because within this framework some
 underlying old nuclear symmetries, namely those implied by
 Wigner and Serber forces, are vividly
 displayed~\cite{CalleCordon:2008cz,CalleCordon:2009ps,RuizArriola:2009bg,Arriola:2010qk}.

In the present paper we want to show that in fact these parameters can
uniquely be determined from known NN scattering threshold parameters
by rather simple calculations by just coarse grain the interaction
over all wavelengths larger than the typical ones occurring in finite
nuclei. As we will show, this
introduces a momentum scale $\Lambda$ in the 9 effective parameters
$t_{0,1,2}$, $x_{0,1,2}$ and $t_{U,T,V}$ which allow to connect the
two body problem to the many body problem.
Going beyond Eq.~(\ref{eq:skyrme2}) requires further information than
just two-body low energy scattering, in particular knowledge about
three and four body forces and their scale dependence consistently
inherited from their NN counterpart.  The finite $k_F$ situation
relevant for heavy nuclei and nuclear matter involves mixing between
operators with different particle number and, in principle, could be
conveniently tackled with the method outlined in
Ref.~\cite{delaPlata:1996fe} where the lack of genuine medium effects
is manifestly built in.

For completeness, we review here the $V_{\rm low k}$
approach~\cite{Bogner:2003wn} in a way that our points can be easily
stated. The starting point is a {\it given} phenomenological NN
potential, $V$, and usually denominated {\it bare potential}, whence
the scattering amplitude or $T$ matrix is obtained as the solution of
the half-off shell Lippmann-Schwinger (LS) coupled channel equation in
the CM system
\begin{eqnarray}
&& T_{l',l}^J (k',k; k^2) = V_{l',l}^J (k',k) 
\nonumber \\ 
&&+ \sum_{l''} \int_0^\infty \, 
\frac{M_N}{(2\pi)^3} 
\frac{dq \, q^2}{k^2-q^2} V_{l',l''}^J  ( k' , q) T_{l'',l}^J ( q , k;k^2) \, , 
\end{eqnarray}
where $J$ is the total angular momentum and $l,l'$ are orbital angular
momentum quantum numbers $p,p',q$ are CM momenta and $M_N$ is the
Nucleon mass. The unitary (coupled channel) S-matrix is obtained as
usual
\begin{eqnarray}
S_{l',l}^J (p) = \delta_{l',l} -  {\rm i} \frac{p M_N}{8 \pi^2} T_{l',l}^J (p,p) \, . 
\end{eqnarray}
Using the matrix representation ${\bf S}^J = ({\bf M}^J - i {\bf 1}
)({\bf M}^J + i {\bf 1})^{-1} $ with $({\bf M}^J)^\dagger = {\bf M}^J$
a hermitian coupled channel matrix, at low energies the effective
range theory for coupled channels reads
\begin{eqnarray}
 p^{l+l'+1} M_{l',l}^J (p) = -(\alpha^{-1})^{J}_{l,l'} + \frac12  (r)^{J}_{l,l'}p^2+  (v)^{J}_{l,l'}p^4 + \dots
\label{eq:ERE-coup}
\end{eqnarray}
which in the absence of mixing  and using  $S_l(p) = e^{2 i
  \delta_l (p)}$ reduces to the well-known expression
\begin{eqnarray}
p^{2 l +1} \cot \delta_l (p) = - \frac{1}{\alpha_l} + \frac12 r_l p^2
+ v_l p^4 + \dots
\label{eq:ERE}
\end{eqnarray}
An extensive study and determination of the low energy parameters for
all partial waves has been carried out in
Ref.~\cite{PavonValderrama:2005ku} for both the NijmII and the Reid93
potentials~\cite{Stoks:1994wp} yielding similar numerical results.
Dropping these coupled channel indices for simplicity the $V_{\rm low
  k}$ potential is then defined by the equation
\begin{eqnarray}
&& T(k',k;k^2) = V_{\rm low k} ( k' , k) \nonumber \\  
&& + \int_0^\Lambda \, \frac{M_N}{(2\pi)^3} \frac{dq \, q^2}{k^2-q^2}
V_{\rm low k} ( k' , q) T ( q , k;k^2) \, , 
\label{eq:Tlowk}
\end{eqnarray}
where $ (k,k') \le \Lambda $. We use here a sharp three-dimensional
cut-off $\Lambda$ to separate between low and high momenta since our
results are not sensitive to the specific form of the regularization.
Thus, eliminating the $T$ matrix we get the equation for the effective
potential which evidently depends on the cut-off scale $\Lambda$ and
corresponds to the effective interaction which nucleons see when all
momenta higher then the momentum scale $\Lambda$ are integrated out.
It has been found~\cite{Bogner:2003wn} that high precision potential
models, i.e. fitting the NN data to high accuracy incorporating One
Pion Exchange (OPE) at large distances and describing the deuteron
form factors, collapse into a unique self-adjoint nonlocal potential
for $\Lambda \sim 400-450 {\rm MeV}$. This is a not a unreasonable
result since all the potentials provide a rather satisfactory
description of elastic NN scattering data up to $p \sim 400 {\rm
  MeV}$. Note that this universality requires a marginal effect of
off-shell ambiguities (beyond OPE off-shellness), which is a great
advantage as this is a traditional source for uncertainties in nuclear
structure.  Actually, in the extreme limit when $\Lambda \to 0$ one is
left with zero energy {\it on shell} scattering yielding $T(k,k) \to
-(2\pi)^3 \alpha_0 /M_N$.

Moreover, for sufficiently small $\Lambda$, the potential which comes
out from eliminating high energy modes can be accurately represented
as the sum of the truncated original potential and a polynomial in the
momentum~\cite{Holt:2003rj}. However, as discussed in
\cite{CalleCordon:2009ps} a more convenient representation is to
separate off all polynomial dependence explicitly from the original
potential
\begin{eqnarray}
V_{\rm low k} ( k' , k) = \bar V_{\rm NN} ( k' , k) + \bar V_{\rm CT}^\Lambda ( k' , k)\, , 
\label{eq:vlowk2} 
\end{eqnarray} 
with $ (k,k') \le \Lambda $, so that if $\bar V_{\rm CT}^\Lambda ( k'
, k)$ contains up to ${\cal O} (p^n)$ then $\bar V_{\rm NN} ( k' , k)$
starts off at ${\cal O} (p^{n+1})$, i.e. the next higher order. This
way the departures from a pure polynomial may be viewed as true and
explicit effects due to the potential and more precisely from the
logarithmic left cut located at CM momentum $p= i m/2$ at the partial
wave amplitude level due to particle exchange with mass
$m$. Specifically,
\begin{eqnarray}
V_{\rm CT}^\Lambda ( k' , k) = k^l k'^{l'} \left[ C_J^{l l'}
  (\Lambda)+ D_J^{l l'}(\Lambda) ( k^2+ k'^2 ) + \dots \right] \, ,
\label{eq:Vlowk} 
\end{eqnarray} 
where the coefficients $C_J^{l l'}(\Lambda) $ and $D_J^{l l'}(\Lambda)
$ include all contributions to the effective interaction at low
energies.  Although we cannot calculate them {\it ab initio} we may
relate them to low energy scattering data, in harmony with the
expectation that off-shell effects are marginal. Not surprisingly the
physics encoding the effective interaction in Eq.~(\ref{eq:Vlowk})
will be related to the threshold parameters defined by
Eq.~(\ref{eq:ERE-coup}). Thus, the relevance of specific microscopic
nuclear forces to the effective (coarse grained) forces has to do with
the extent to which these threshold parameters are described by the
underlying forces and not so much with their detailed structure.  We
will discuss below the limitations to this universal pattern.


Using the partial wave projection~\cite{1971NuPhA.176..413E} we get
the potentials in different angular momentum channels.  These
parameters can be related to the spectroscopic notation used in
Ref.~\cite{Epelbaum:1999dj}. The S- and P-wave potentials are
\begin{eqnarray}
V_{^1S_0} (p',p) &=& C_{^1S_0} + D_{^1S_0} (p'^2+p^2) \, ,\nonumber \\ 
V_{^3S_1} (p',p) &=& C_{^3S_1} + D_{^3S_1} (p'^2+p^2) \, ,\nonumber \\ 
V_{E_1} (p',p) &=& D_{E_1} p^2 \, , \nonumber \\ 
V_{^3P_0} (p',p) &=& C_{^3P_0} p' p \, ,  \nonumber \\ 
V_{^3P_1} (p',p) &=& C_{^3P_1} p' p \, , \nonumber \\ 
V_{^3P_2} (p',p) &=& C_{^3P_2} p' p \, , \nonumber \\ 
V_{^1P_1} (p',p) &=& C_{^1P_1} p' p \, .  
\label{eq:Vlowkp2}
\end{eqnarray}
The 9 effective parameters depend on the scale $\Lambda$ and can
be related to the effective force representation $t_{0,1,2}$,
$x_{0,1,2}$ and $t_{V,U,S}$ of Eq.~(\ref{eq:skyrme2}) by the following
explicit relations,
\begin{eqnarray}
t_0 &=& \frac1{8\pi}\left( C_{^3S_1}+C_{^1S_0} \right) \, ,\nonumber
\\
 x_0 &=& \frac{C_{^3S_1}-C_{^1S_0}}{C_{^3S_1}+C_{^1S_0}} \, , \nonumber
\\
 t_1 &=& \frac1{8\pi}\left( D_{^3S_1}+D_{^1S_0} \right) \, , \nonumber
\\
 x_1 &=& \frac{D_{^3S_1}-D_{^1S_0}}{D_{^3S_1}+D_{^1S_0}} \, , \nonumber 
\\
 t_2 &=& \frac1{32\pi}\left( 9 C_{^1P_1}+C_{^3P_0}+ 3 C_{^3P_1}+5
C_{^3P_2} \right) \, , \nonumber 
\\ 
x_2 &=& \frac{ -9 C_{^1P_1}+C_{^3P_0}+ 3 C_{^3P_1}+5
C_{^3P_2}}{ 9 C_{^1P_1}+C_{^3P_0}+ 3 C_{^3P_1}+5
C_{^3P_2}} \, , \nonumber \\ 
t_T &=& -\frac3{4 \sqrt{2}\pi} D_{E_1} \, , \nonumber \\ 
t_V &=& \frac1{32\pi} 
\left( 2 C_{^3P_0}+ 3 C_{^3P_1}-5 C_{^3P_2} \right) \, ,
\nonumber
\\
t_U &=& \frac1{16\pi} 
\left( -2 C_{^3P_0}+ 3 C_{^3P_1}- C_{^3P_2} \right) \, . 
\end{eqnarray}


\begin{figure}[ttt]
\includegraphics[height=5.5cm,width=7cm,angle=0]{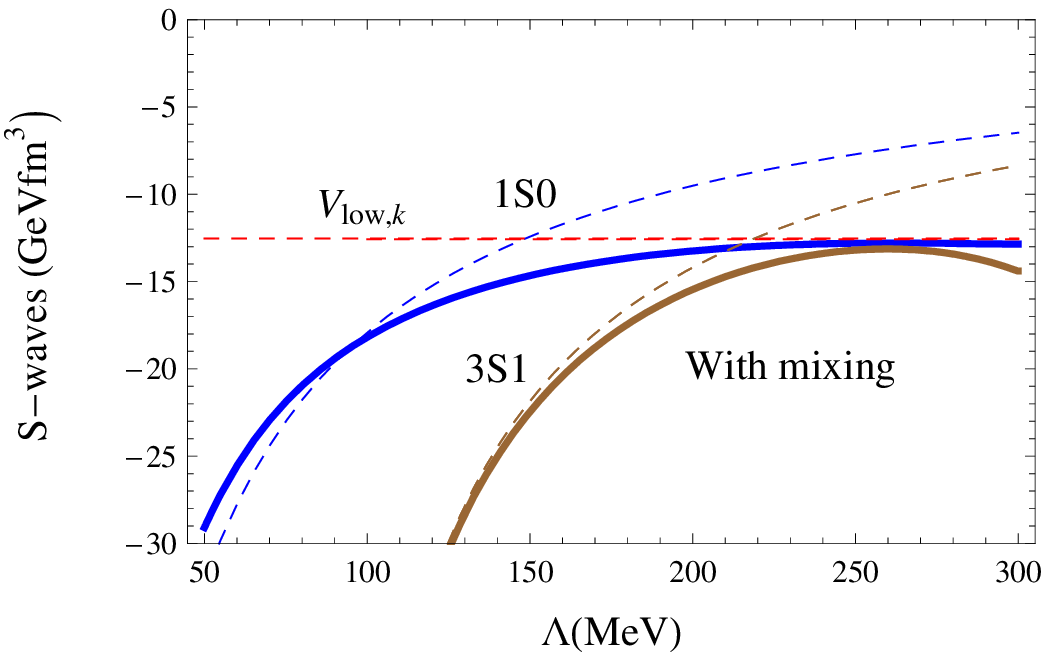}
\hskip.5cm 
\includegraphics[height=5.5cm,width=7cm,angle=0]{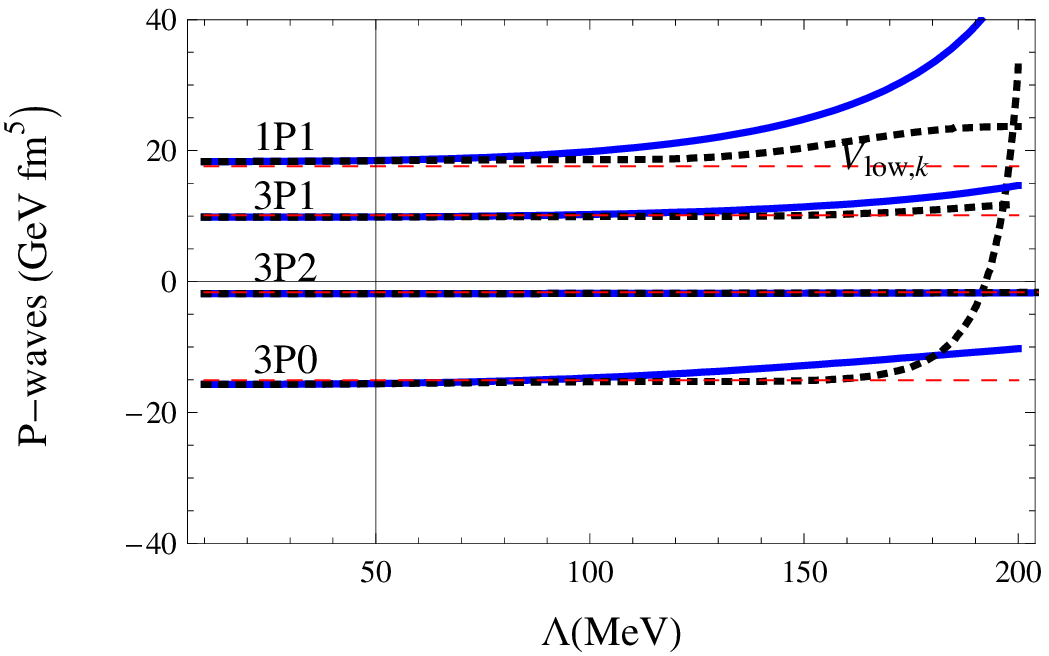}
\caption{(Color on-line) Counterterms for the S- (in ${\rm MeV} {\rm
    fm}^3$, upper panel) and P-waves (${\rm MeV} {\rm fm}^5$, lower
  panel) as a function of the momentum scale $\Lambda$ (in MeV).  C's
  from Eqs.~(\ref{eq:Vlowkp2}) solving Eqs.~(\ref{eq:Tlowk}) including
  the D's using just the low energy threshold parameters from
  Ref.~\cite{PavonValderrama:2005ku} (Thick Solid).  C's extracted
  from the diagonal $V_{\rm low k} (p,p)$
  potentials~\cite{Bogner:2003wn} at fixed $\Lambda= 420 {\rm MeV}$
  for the Argonne-V18~\cite{Wiringa:1994wb} (dashed). C's for P-waves
  including D-terms without mixings (thick dotted).}
\label{fig:vlowk}
\end{figure}

\begin{figure}[ttt]
\includegraphics[height=5.5cm,width=7cm,angle=0]{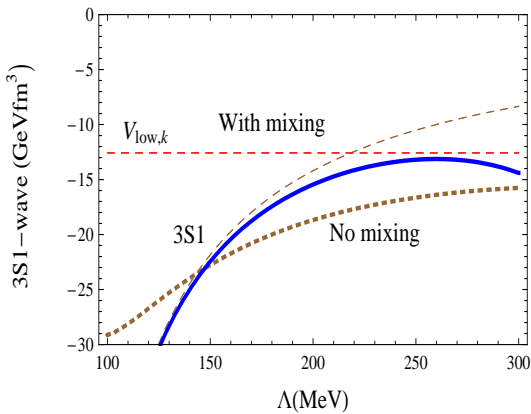}
\caption{(Color on-line) $S-D$ waves mixing on $C_{^3S_1}$ (in ${\rm
    MeV}{\rm fm}^3$) as a function of the momentum space cut-off
  $\Lambda$ (in MeV) in the $^3S_1$ channel. We compare when including
  a) only $C_{^3S_1}$ (dashed), b) $C_{^3S_1}$ and $D_{^3S_1}$
  (dotted), c) $C_{^3S_1}$, $D_{^3S_1}$ and $D_{E_1}$ (solid).  Wigner
  symmetry is displayed when the mixing is included. See also
  Fig.~\ref{fig:vlowk} and Eq.~(\ref{eq:vlowk2}) in the main text.}
\label{fig:nomix-mix}
\end{figure}

\begin{figure*}[tbc]
\includegraphics[height=3.5cm,width=4.4cm,angle=0]{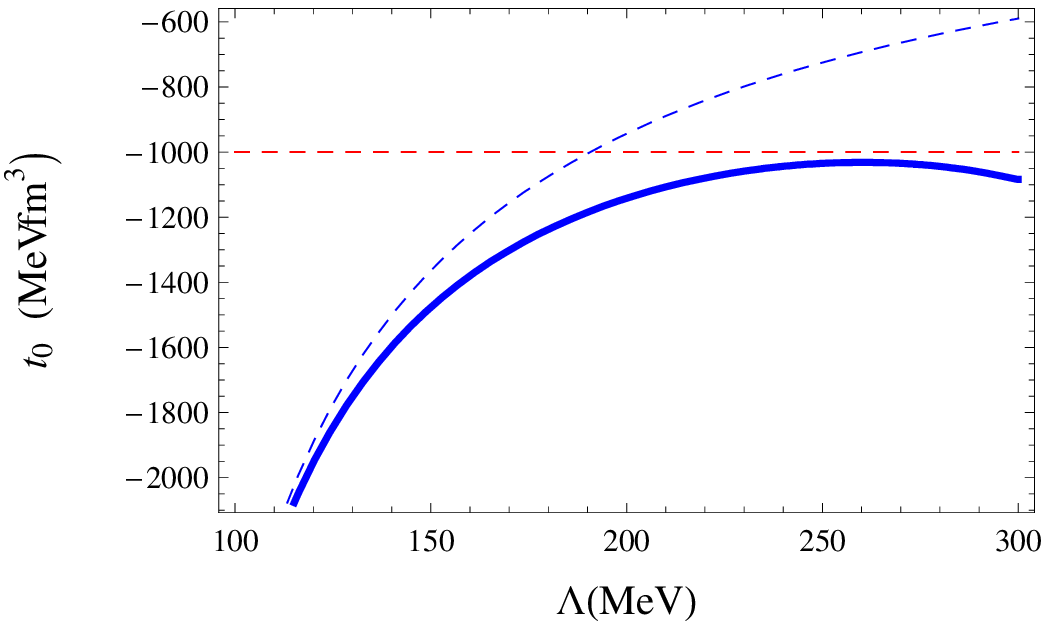}
\hskip.3cm 
\includegraphics[height=3.5cm,width=4.4cm,angle=0]{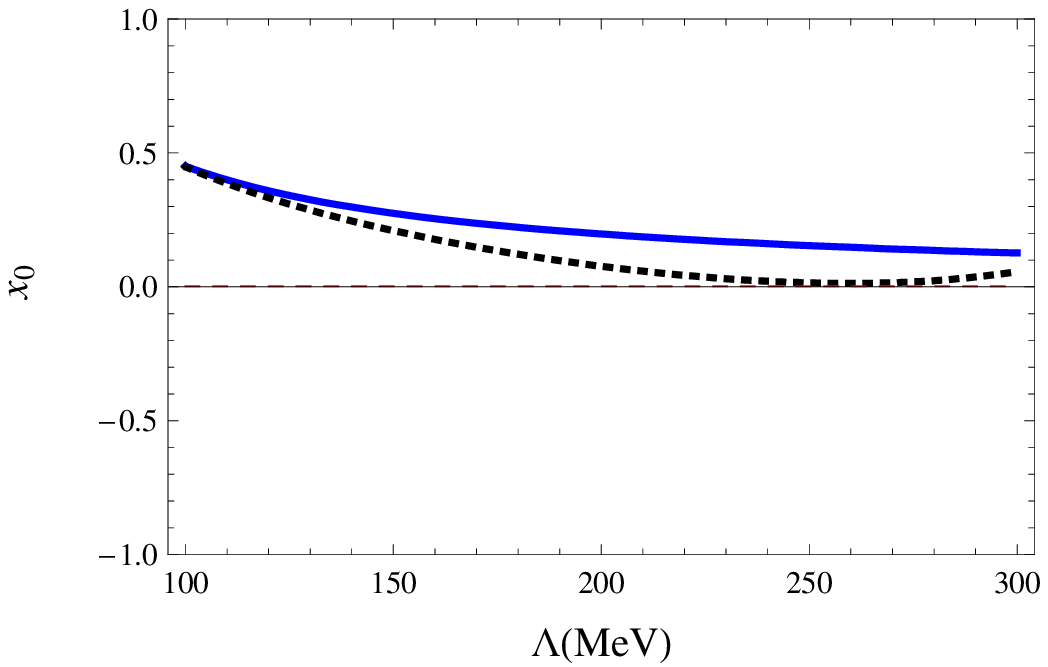}
\hskip.3cm 
\includegraphics[height=3.5cm,width=4.4cm,angle=0]{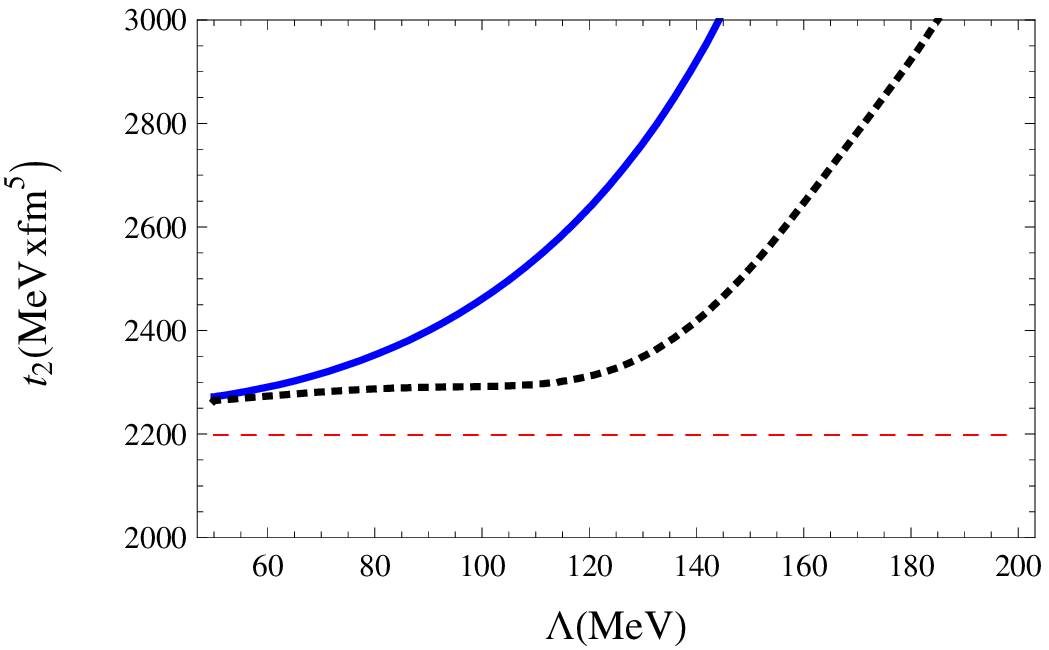}\\
\includegraphics[height=3.5cm,width=4.4cm,angle=0]{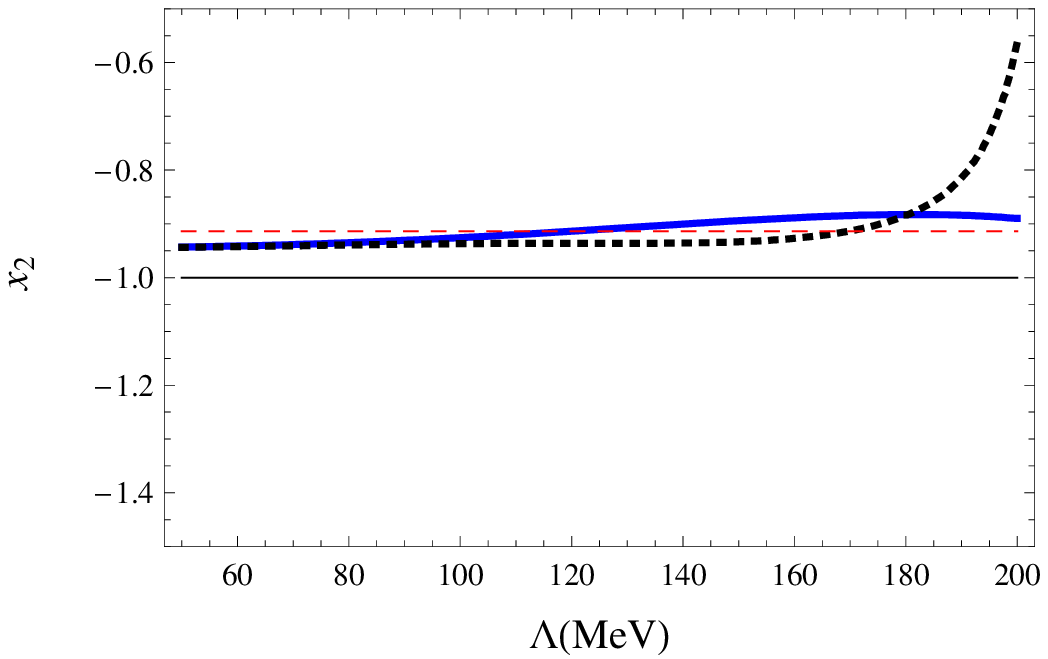}
\hskip.3cm 
\includegraphics[height=3.5cm,width=4.4cm,angle=0]{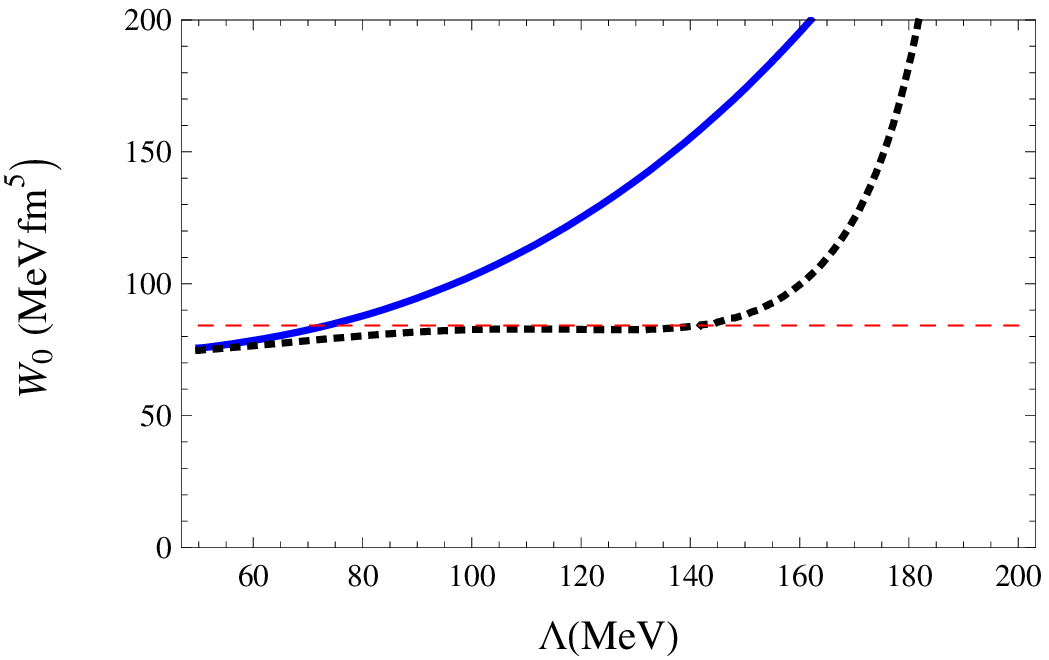}
\hskip.3cm 
\includegraphics[height=3.5cm,width=4.4cm,angle=0]{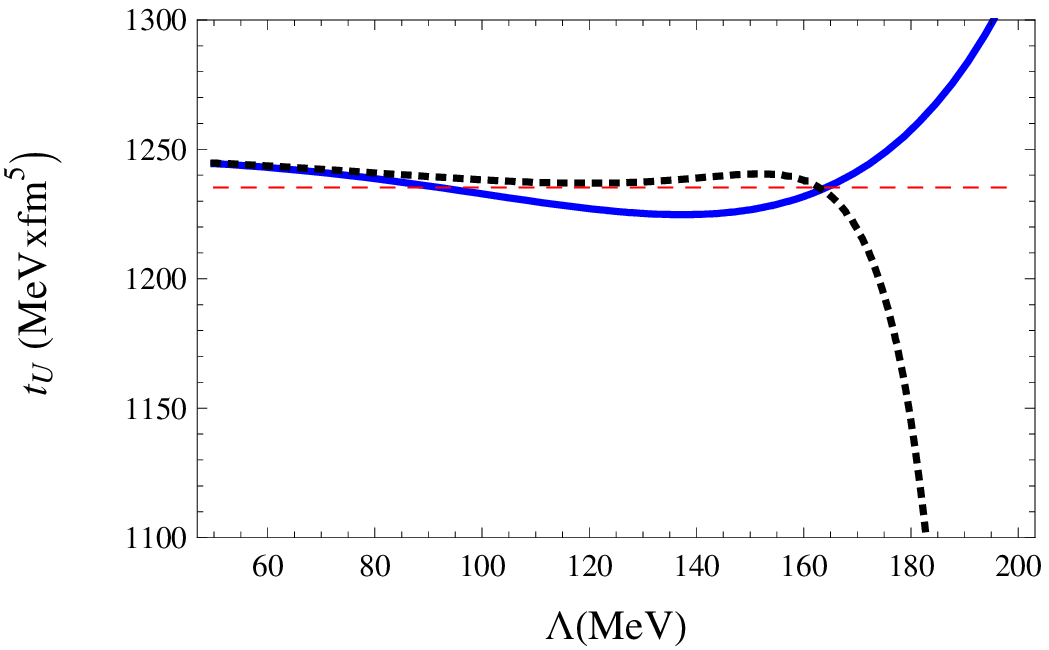}
\caption{(Color on line) Skyrme force parameters as a function of the
  scale $\Lambda$ (in MeV). We compare with the
  Argonne-V18~\cite{Wiringa:1994wb} exact $V_{\rm low k}$ values
  evaluated at $\Lambda=420 {\rm MeV}$~\cite{Bogner:2003wn}. See also
  Fig.~\ref{fig:vlowk} and main text.  }
\label{fig:skyrme}
\end{figure*}


The corresponding T-matrices are conveniently solved by factoring out
the centrifugal terms
\begin{eqnarray}
T_{l',l}(k',k) =  k^l k'^{l'} \left[ t_{l',l}^J (p) + u_{l',l}(p) (k^2+k'^2) + \dots
\right]  
\end{eqnarray}
which reduce the LS equation to a finite set of algebraic
equations which are analytically solvable (see
e.g. Ref.~\cite{Entem:2007jg} and references therein). In the simplest
case where only the $C's$ are taken into account the explicit
solutions for S- and P-waves are,
\begin{eqnarray}
C_{S} (\Lambda) &=& \frac{16 \pi^2 \alpha_{0}}{M_N (1 - 2 \alpha_{0}
  \Lambda/\pi)} \, , \nonumber \\ C_P (\Lambda) &=& \frac{16 \pi^2
  \alpha_1}{M_N (1 - 2 \alpha_{1} \Lambda^3/ 3 \pi)  } \, , 
\label{eq:C's}
\end{eqnarray}
where $\alpha_0$($\alpha_1$) is the scattering length (volume) defined by
Eq.~(\ref{eq:ERE}). The Eq.~(\ref{eq:C's}) illustrates the difference
between a Fermi pseudo-potential and a coarse grained potential as the
former corresponds to $\alpha_0 \Lambda \ll 1 $ where $ C_{S}
(\Lambda) \sim 16 \pi^2 \alpha_{0}/M_N $. In the case $\alpha_0 \Lambda
\gg 1 $ one has instead $ C_{S} (\Lambda) \sim - 8 \pi / ( M
\Lambda)$. Full solutions including the $D$'s are also analytical
although a bit messier, so we do not display them explicitly.  
They rely on Eq.~(\ref{eq:ERE-coup}) with $\alpha_{^1S_0}$,
$\alpha_{^3S_1}$, $\alpha_{^3P_0}$, $\alpha_{^3P_1}$,
$\alpha_{^3P_2}$, $\alpha_{^1P_1}$, $\alpha_{E_1}$, $r_{^3S_1}$ and
$r_{^1S_0}$ (see Ref.~\cite{PavonValderrama:2005ku} for numerical
values for NijmII and Reid93 potentials).  At the order considered
here we just mention that while all P-waves constants run
independently of each other with $\Lambda$ the spin-singlet parameters
$C_{^1S_0} $, $D_{^1S_0} $ on the one hand and the spin-triplet
parameters $C_{^3S_1} $, $D_{^3S_1} $ and $C_{E_1} $ on the other are
intertwined.

We now turn to our numerical results.  As can be seen from
Fig.~\ref{fig:vlowk} the comparison of contact interactions using
threshold parameters with $V_{\rm low k}$ results evolved to
$\Lambda=420 {\rm MeV}$~\cite{Bogner:2003wn} (note the different
normalization as ours) from the Argonne-V18 bare
potential~\cite{Wiringa:1994wb} are saturated for $\Lambda=250 {\rm
  MeV}$ for S-waves and for much lower cut-offs for P-waves. Note that
this holds regardless on the details of the potential as we only need
the low energy threshold parameters as determined e.g. in
Ref.~\cite{PavonValderrama:2005ku}. The strong dependence observed at
larger $\Lambda$ values just reflects the inadequacy of the second
order truncation in Eqs.~(\ref{eq:Vlowkp2}). This also reflects in the
$25\%-50\%$ inaccuracy are off the exact $V_{\rm low k}$ of the D's
themselves despite showing plateaus, and thus will not be discussed
any further.

The identity $C_{^1S_0}(\Lambda)=C_{^3S_1}(\Lambda)$ for $\Lambda \ge
250 {\rm MeV}$ features the appearance of Wigner symmetry as pointed
out in Ref.~\cite{CalleCordon:2008cz}, but now we see that this does
not depend on details of the force. Actually, the effect of the
$^3S_1-^3D_1$wave mixing represented by a non-vanishing off diagonal
potential $V_{E_1} (p',p)$ becomes essential to achieve this identity
(a fact disregarded in Ref.~\cite{Mehen:1999qs}). As can be seen from
Fig.~\ref{fig:nomix-mix} there is a large mismatch at values of
$\Lambda \sim 200-300 {\rm MeV}$ when $D_{E_1}$ is set to zero (and
hence $\alpha_{E_1}=0$) as compared with the case $D_{E_1} \neq 0$.

The scale dependence of the Skyrme interaction parameters (not
involving the D's) can be seen in Fig.~\ref{fig:skyrme} in comparison
with the $V_{\rm low k} $ potentials~\cite{Bogner:2003wn} deduced from
the Argonne-V18 bare potentials~\cite{Wiringa:1994wb}. The plateaus
observed in the different partial waves are corroborated here as well
as a remarkable accuracy in reproducing the exact $V_{\rm low k} $
numbers. Moreover, 
the weak cut-off dependence of the spin orbit interaction
observed in Fig.~\ref{fig:skyrme} suggests taking $\Lambda \to 0$ in
which case
\begin{eqnarray}
W_0 = \frac{\pi}{ 2 M_N} \left( 2 \alpha_{^3P_0} + 3 \alpha_{^3P_1} -5
\alpha_{^3P_2} \right) \, , 
\end{eqnarray}
which upon using Ref.~\cite{PavonValderrama:2005ku} yields $ W_0 = 72
{\rm MeV fm}^4$. This numerical value reproduces within less than
$10\%$ the exact $V_{\rm low k}$ value. As can be seen from
Fig.~\ref{fig:skyrme} the effective range correction $r_1$ provides
via additional $D$ coefficients  the missing
contribution. This is a bit lower than what it is found in
phenomenological approaches from the $p_{3/2}-p_{1/2}$ level splitting
in $^{16}$O~\cite{Bender:2003jk}. In any case, the comparison with
phenomenological approaches based on mean field calculations may be
tricky since as already mentioned not all the terms are always kept,
and selective fits to finite nuclear properties may overemphasize the
role played by specific terms.

It has recently been argued that counterterms are fingerprints of long
distance
symmetries~\cite{CalleCordon:2008cz,CalleCordon:2009ps,RuizArriola:2009bg}.
This remarkable result holds regardless on the nature of the forces
and applies in particular to both Wigner and Serber symmetries.  We
confirm that to great accuracy, $x_0=0$ (Wigner symmetry) and
$x_2=-1$ (Serber symmetry). The astonishing large-$N_c$ ($N_c$ is the
number of colours in QCD) relations discussed in
Refs.~\cite{CalleCordon:2008cz,CalleCordon:2009ps,RuizArriola:2009bg,Arriola:2010qk}
provide a direct link to the underlying quark and gluon dynamics and
after~\cite{Kaplan:1996rk} suggests a $1/N_c^2$ accuracy of the Wigner
symmetry in even-L partial waves.  Wigner symmetry has proven crucial
in Nuclear coarse lattice ($a \sim 2{\rm fm}$)
calculations~\cite{Lee:2008fa} in sidestepping the sign problem for
fermions. As we see for the scales typically involved there this works
with great accuracy already at $\Lambda \sim 250 {\rm MeV}$.  Taking
into account that we deal with low energies, it is thus puzzling that
Chiral interactions to N$^3$LO~\cite{Entem:2003ft} having chiral
cut-offs $\Lambda_\chi \sim 600 {\rm MeV} $ tend to violate Wigner
symmetry in the $V_{\rm low k}$ sense, i.e. $C_{^1S_0}^\chi \neq
C_{^3S_1}^\chi $, whereas smaller values $\Lambda_\chi \sim 450 {\rm
  MeV} $~\cite{CalleCordon:2009ps} are preferred.

Within the low energy expansion we have neglected terms ${\cal O}
({\bf p'}^4 , {\bf p}^4 , {\bf p'}^2 {\bf p}^2 )$ which correspond to
P-waves and S-wave range corrections. In configuration space this
corresponds to a dimensional expansion, since $\delta (\vec r_{12}) =
{\cal O} (\Lambda^3) $ and $ \{P^2, \delta (\vec r_{12}) \} = {\cal O}
(\Lambda^5) $, $ \{P^4, \delta (\vec r_{12}) \} = {\cal O} (\Lambda^7)
$. Within such a scheme going to higher orders requires also to
include three-body interactions, $\sim \delta (r_{12}) \delta (r_{13})
= {\cal O} (\Lambda^6) $. Actually, at the two body level there are
more potential parameters than low energy threshold parameters. For
instance, in the $^1S_0$ channel one has two independent hermitean
operators, ${\bf p'}^4 + {\bf p}^4 $ and $ 2 {\bf p'}^2 {\bf p}^2 $
(which are on-shell equivalent), but only one $v_{^1S_0}$ threshold
parameter in the low energy expansion (see
Eq.~(\ref{eq:ERE-coup})). As it was shown in Ref.~\cite{Amghar:1995av}
(see also Ref.~\cite{Furnstahl:2000we}) these two features are
interrelated since this two body off-shell ambiguity is cancelled when
a three body observable, like e.g.  the triton binding energy, is
fixed .  An intriguing aspect of the present investigation is the
modification induced by potential tails due to e.g. pion exchange
which cannot be represented by a polynomial since particle exchange
generates a cut in the complex energy plane. The important issue,
however, is that the low scale saturation unveiled in the present
paper works accurately just to second order as long as the low energy
parameters determined from on-shell scattering are properly
reproduced.

{\it I thank M. Pav\'on Valderrama and  L.L. Salcedo
for a critical reading of the ms
  and Jes\'us Navarro, A. Calle Cord\'on, T. Frederico
  and V.S. Timoteo for discussions. Work supported by the Spanish DGI
  and FEDER funds with grant FIS2008-01143/FIS, Junta de
  Andaluc{\'\i}a grant FQM225-05.}


\begin{thebibliography}{36}
\expandafter\ifx\csname natexlab\endcsname\relax\def\natexlab#1{#1}\fi
\expandafter\ifx\csname bibnamefont\endcsname\relax
  \def\bibnamefont#1{#1}\fi
\expandafter\ifx\csname bibfnamefont\endcsname\relax
  \def\bibfnamefont#1{#1}\fi
\expandafter\ifx\csname citenamefont\endcsname\relax
  \def\citenamefont#1{#1}\fi
\expandafter\ifx\csname url\endcsname\relax
  \def\url#1{\texttt{#1}}\fi
\expandafter\ifx\csname urlprefix\endcsname\relax\def\urlprefix{URL }\fi
\providecommand{\bibinfo}[2]{#2}
\providecommand{\eprint}[2][]{\url{#2}}

\bibitem[{\citenamefont{Moshinsky}(1958)}]{Moshinsky195819}
\bibinfo{author}{\bibfnamefont{M.}~\bibnamefont{Moshinsky}},
  \bibinfo{journal}{Nuclear Physics} \textbf{\bibinfo{volume}{8}},
  \bibinfo{pages}{19 } (\bibinfo{year}{1958}).

\bibitem[{\citenamefont{Skyrme}(1959)}]{Skyrme:1959zz}
\bibinfo{author}{\bibfnamefont{T.}~\bibnamefont{Skyrme}},
  \bibinfo{journal}{Nucl. Phys.} \textbf{\bibinfo{volume}{9}},
  \bibinfo{pages}{615} (\bibinfo{year}{1959}).

\bibitem[{\citenamefont{Vautherin and Brink}(1972)}]{Vautherin:1971aw}
\bibinfo{author}{\bibfnamefont{D.}~\bibnamefont{Vautherin}} \bibnamefont{and}
  \bibinfo{author}{\bibfnamefont{D.~M.} \bibnamefont{Brink}},
  \bibinfo{journal}{Phys. Rev.} \textbf{\bibinfo{volume}{C5}},
  \bibinfo{pages}{626} (\bibinfo{year}{1972}).

\bibitem[{\citenamefont{Negele and Vautherin}(1972)}]{Negele:1972zp}
\bibinfo{author}{\bibfnamefont{J.~W.} \bibnamefont{Negele}} \bibnamefont{and}
  \bibinfo{author}{\bibfnamefont{D.}~\bibnamefont{Vautherin}},
  \bibinfo{journal}{Phys. Rev.} \textbf{\bibinfo{volume}{C5}},
  \bibinfo{pages}{1472} (\bibinfo{year}{1972}).

\bibitem[{\citenamefont{Chabanat et~al.}(1997)\citenamefont{Chabanat, Meyer,
  Bonche, Schaeffer, and Haensel}}]{Chabanat:1997qh}
\bibinfo{author}{\bibfnamefont{E.}~\bibnamefont{Chabanat}},
  \bibinfo{author}{\bibfnamefont{J.}~\bibnamefont{Meyer}},
  \bibinfo{author}{\bibfnamefont{P.}~\bibnamefont{Bonche}},
  \bibinfo{author}{\bibfnamefont{R.}~\bibnamefont{Schaeffer}},
  \bibnamefont{and} \bibinfo{author}{\bibfnamefont{P.}~\bibnamefont{Haensel}},
  \bibinfo{journal}{Nucl. Phys.} \textbf{\bibinfo{volume}{A627}},
  \bibinfo{pages}{710} (\bibinfo{year}{1997}).

\bibitem[{\citenamefont{Bender et~al.}(2003)\citenamefont{Bender, Heenen, and
  Reinhard}}]{Bender:2003jk}
\bibinfo{author}{\bibfnamefont{M.}~\bibnamefont{Bender}},
  \bibinfo{author}{\bibfnamefont{P.-H.} \bibnamefont{Heenen}},
  \bibnamefont{and} \bibinfo{author}{\bibfnamefont{P.-G.}
  \bibnamefont{Reinhard}}, \bibinfo{journal}{Rev. Mod. PHys.}
  \textbf{\bibinfo{volume}{75}}, \bibinfo{pages}{121} (\bibinfo{year}{2003}).

\bibitem[{\citenamefont{{Moszkowski} and {Scott}}(1960)}]{1960AnPhy..11...65M}
\bibinfo{author}{\bibfnamefont{S.~A.} \bibnamefont{{Moszkowski}}}
  \bibnamefont{and} \bibinfo{author}{\bibfnamefont{B.~L.}
  \bibnamefont{{Scott}}}, \bibinfo{journal}{Annals of Physics}
  \textbf{\bibinfo{volume}{11}}, \bibinfo{pages}{65} (\bibinfo{year}{1960}).

\bibitem[{\citenamefont{Dean et~al.}(2004)\citenamefont{Dean, Engeland,
  Hjorth-Jensen, Kartamyshev, and Osnes}}]{Dean:2004ck}
\bibinfo{author}{\bibfnamefont{D.~J.} \bibnamefont{Dean}},
  \bibinfo{author}{\bibfnamefont{T.}~\bibnamefont{Engeland}},
  \bibinfo{author}{\bibfnamefont{M.}~\bibnamefont{Hjorth-Jensen}},
  \bibinfo{author}{\bibfnamefont{M.}~\bibnamefont{Kartamyshev}},
  \bibnamefont{and} \bibinfo{author}{\bibfnamefont{E.}~\bibnamefont{Osnes}},
  \bibinfo{journal}{Prog. Part. Nucl. Phys.} \textbf{\bibinfo{volume}{53}},
  \bibinfo{pages}{419} (\bibinfo{year}{2004}), \eprint{nucl-th/0405034}.

\bibitem[{\citenamefont{Coraggio et~al.}(2009)\citenamefont{Coraggio, Covello,
  Gargano, Itaco, and Kuo}}]{Coraggio:2008in}
\bibinfo{author}{\bibfnamefont{L.}~\bibnamefont{Coraggio}},
  \bibinfo{author}{\bibfnamefont{A.}~\bibnamefont{Covello}},
  \bibinfo{author}{\bibfnamefont{A.}~\bibnamefont{Gargano}},
  \bibinfo{author}{\bibfnamefont{N.}~\bibnamefont{Itaco}}, \bibnamefont{and}
  \bibinfo{author}{\bibfnamefont{T.~T.~S.} \bibnamefont{Kuo}},
  \bibinfo{journal}{Prog. Part. Nucl. Phys.} \textbf{\bibinfo{volume}{62}},
  \bibinfo{pages}{135} (\bibinfo{year}{2009}), \eprint{0809.2144}.

\bibitem[{\citenamefont{Friedrich and Reinhard}(1986)}]{Friedrich:1986zza}
\bibinfo{author}{\bibfnamefont{J.}~\bibnamefont{Friedrich}} \bibnamefont{and}
  \bibinfo{author}{\bibfnamefont{P.~G.} \bibnamefont{Reinhard}},
  \bibinfo{journal}{Phys. Rev.} \textbf{\bibinfo{volume}{C33}},
  \bibinfo{pages}{335} (\bibinfo{year}{1986}).

\bibitem[{\citenamefont{Klupfel et~al.}(2009)\citenamefont{Klupfel, Reinhard,
  Burvenich, and Maruhn}}]{Klupfel:2008af}
\bibinfo{author}{\bibfnamefont{P.}~\bibnamefont{Klupfel}},
  \bibinfo{author}{\bibfnamefont{P.~G.} \bibnamefont{Reinhard}},
  \bibinfo{author}{\bibfnamefont{T.~J.} \bibnamefont{Burvenich}},
  \bibnamefont{and} \bibinfo{author}{\bibfnamefont{J.~A.}
  \bibnamefont{Maruhn}}, \bibinfo{journal}{Phys. Rev.}
  \textbf{\bibinfo{volume}{C79}}, \bibinfo{pages}{034310}
  (\bibinfo{year}{2009}), \eprint{0804.3385}.

\bibitem[{\citenamefont{Furnstahl and Hackworth}(1997)}]{Furnstahl:1997hq}
\bibinfo{author}{\bibfnamefont{R.~J.} \bibnamefont{Furnstahl}}
  \bibnamefont{and} \bibinfo{author}{\bibfnamefont{J.~C.}
  \bibnamefont{Hackworth}}, \bibinfo{journal}{Phys. Rev.}
  \textbf{\bibinfo{volume}{C56}}, \bibinfo{pages}{2875} (\bibinfo{year}{1997}),
  \eprint{nucl-th/9708018}.

\bibitem[{\citenamefont{Kortelainen et~al.}(2010)\citenamefont{Kortelainen,
  Furnstahl, Nazarewicz, and Stoitsov}}]{Kortelainen:2010dt}
\bibinfo{author}{\bibfnamefont{M.}~\bibnamefont{Kortelainen}},
  \bibinfo{author}{\bibfnamefont{R.~J.} \bibnamefont{Furnstahl}},
  \bibinfo{author}{\bibfnamefont{W.}~\bibnamefont{Nazarewicz}},
  \bibnamefont{and} \bibinfo{author}{\bibfnamefont{M.~V.}
  \bibnamefont{Stoitsov}} (\bibinfo{year}{2010}), \eprint{1005.2552}.

\bibitem[{\citenamefont{Baldo et~al.}(2010)\citenamefont{Baldo, Robledo,
  Schuck, and Vinas}}]{Baldo:2010nx}
\bibinfo{author}{\bibfnamefont{M.}~\bibnamefont{Baldo}},
  \bibinfo{author}{\bibfnamefont{L.}~\bibnamefont{Robledo}},
  \bibinfo{author}{\bibfnamefont{P.}~\bibnamefont{Schuck}}, \bibnamefont{and}
  \bibinfo{author}{\bibfnamefont{X.}~\bibnamefont{Vinas}}, \bibinfo{journal}{J.
  Phys.} \textbf{\bibinfo{volume}{G37}}, \bibinfo{pages}{064015}
  (\bibinfo{year}{2010}), \eprint{1005.1810}.

\bibitem[{\citenamefont{Stoitsov et~al.}(2010)}]{Stoitsov:2010ha}
\bibinfo{author}{\bibfnamefont{M.}~\bibnamefont{Stoitsov}} \bibnamefont{et~al.}
  (\bibinfo{year}{2010}), \eprint{1009.3452}.

\bibitem[{\citenamefont{Bogner et~al.}(2003{\natexlab{a}})\citenamefont{Bogner,
  Kuo, Schwenk, Entem, and Machleidt}}]{Bogner:2001gq}
\bibinfo{author}{\bibfnamefont{S.~K.} \bibnamefont{Bogner}},
  \bibinfo{author}{\bibfnamefont{T.~T.~S.} \bibnamefont{Kuo}},
  \bibinfo{author}{\bibfnamefont{A.}~\bibnamefont{Schwenk}},
  \bibinfo{author}{\bibfnamefont{D.~R.} \bibnamefont{Entem}}, \bibnamefont{and}
  \bibinfo{author}{\bibfnamefont{R.}~\bibnamefont{Machleidt}},
  \bibinfo{journal}{Phys. Lett.} \textbf{\bibinfo{volume}{B576}},
  \bibinfo{pages}{265} (\bibinfo{year}{2003}{\natexlab{a}}),
  \eprint{nucl-th/0108041}.

\bibitem[{\citenamefont{Bogner et~al.}(2010)\citenamefont{Bogner, Furnstahl,
  and Schwenk}}]{Bogner:2009bt}
\bibinfo{author}{\bibfnamefont{S.~K.} \bibnamefont{Bogner}},
  \bibinfo{author}{\bibfnamefont{R.~J.} \bibnamefont{Furnstahl}},
  \bibnamefont{and} \bibinfo{author}{\bibfnamefont{A.}~\bibnamefont{Schwenk}},
  \bibinfo{journal}{Prog. Part. Nucl. Phys.} \textbf{\bibinfo{volume}{65}},
  \bibinfo{pages}{94} (\bibinfo{year}{2010}), \eprint{0912.3688}.

\bibitem[{\citenamefont{Lee}(2009)}]{Lee:2008fa}
\bibinfo{author}{\bibfnamefont{D.}~\bibnamefont{Lee}}, \bibinfo{journal}{Prog.
  Part. Nucl. Phys.} \textbf{\bibinfo{volume}{63}}, \bibinfo{pages}{117}
  (\bibinfo{year}{2009}), \eprint{0804.3501}.

\bibitem[{\citenamefont{Calle~Cordon and
  Ruiz~Arriola}(2008)}]{CalleCordon:2008cz}
\bibinfo{author}{\bibfnamefont{A.}~\bibnamefont{Calle~Cordon}}
  \bibnamefont{and}
  \bibinfo{author}{\bibfnamefont{E.}~\bibnamefont{Ruiz~Arriola}},
  \bibinfo{journal}{Phys. Rev.} \textbf{\bibinfo{volume}{C78}},
  \bibinfo{pages}{054002} (\bibinfo{year}{2008}), \eprint{0807.2918}.

\bibitem[{\citenamefont{Calle~Cordon and
  Ruiz~Arriola}(2009)}]{CalleCordon:2009ps}
\bibinfo{author}{\bibfnamefont{A.}~\bibnamefont{Calle~Cordon}}
  \bibnamefont{and}
  \bibinfo{author}{\bibfnamefont{E.}~\bibnamefont{Ruiz~Arriola}},
  \bibinfo{journal}{Phys. Rev.} \textbf{\bibinfo{volume}{C80}},
  \bibinfo{pages}{014002} (\bibinfo{year}{2009}), \eprint{0904.0421}.

\bibitem[{\citenamefont{Ruiz~Arriola and
  Calle~Cordon}(2009)}]{RuizArriola:2009bg}
\bibinfo{author}{\bibfnamefont{E.}~\bibnamefont{Ruiz~Arriola}}
  \bibnamefont{and}
  \bibinfo{author}{\bibfnamefont{A.}~\bibnamefont{Calle~Cordon}},
  \bibinfo{journal}{PoS} \textbf{\bibinfo{volume}{EFT09}}, \bibinfo{pages}{046}
  (\bibinfo{year}{2009}), \eprint{0904.4132}.

\bibitem[{\citenamefont{Arriola and Cordon}(2010)}]{Arriola:2010qk}
\bibinfo{author}{\bibfnamefont{E.~R.} \bibnamefont{Arriola}} \bibnamefont{and}
  \bibinfo{author}{\bibfnamefont{A.~C.} \bibnamefont{Cordon}}
  (\bibinfo{year}{2010}), \eprint{1009.3149}.

\bibitem[{\citenamefont{de~la Plata and Salcedo}(1998)}]{delaPlata:1996fe}
\bibinfo{author}{\bibfnamefont{M.~J.} \bibnamefont{de~la Plata}}
  \bibnamefont{and} \bibinfo{author}{\bibfnamefont{L.~L.}
  \bibnamefont{Salcedo}}, \bibinfo{journal}{J. Phys.}
  \textbf{\bibinfo{volume}{A31}}, \bibinfo{pages}{4021} (\bibinfo{year}{1998}),
  \eprint{hep-th/9609103}.

\bibitem[{\citenamefont{Bogner et~al.}(2003{\natexlab{b}})\citenamefont{Bogner,
  Kuo, and Schwenk}}]{Bogner:2003wn}
\bibinfo{author}{\bibfnamefont{S.~K.} \bibnamefont{Bogner}},
  \bibinfo{author}{\bibfnamefont{T.~T.~S.} \bibnamefont{Kuo}},
  \bibnamefont{and} \bibinfo{author}{\bibfnamefont{A.}~\bibnamefont{Schwenk}},
  \bibinfo{journal}{Phys. Rept.} \textbf{\bibinfo{volume}{386}},
  \bibinfo{pages}{1} (\bibinfo{year}{2003}{\natexlab{b}}),
  \eprint{nucl-th/0305035}.

\bibitem[{\citenamefont{Pavon~Valderrama and
  Arriola}(2005)}]{PavonValderrama:2005ku}
\bibinfo{author}{\bibfnamefont{M.}~\bibnamefont{Pavon~Valderrama}}
  \bibnamefont{and} \bibinfo{author}{\bibfnamefont{E.~R.}
  \bibnamefont{Arriola}}, \bibinfo{journal}{Phys. Rev.}
  \textbf{\bibinfo{volume}{C72}}, \bibinfo{pages}{044007}
  (\bibinfo{year}{2005}).

\bibitem[{\citenamefont{Stoks et~al.}(1994)\citenamefont{Stoks, Klomp,
  Terheggen, and de~Swart}}]{Stoks:1994wp}
\bibinfo{author}{\bibfnamefont{V.~G.~J.} \bibnamefont{Stoks}},
  \bibinfo{author}{\bibfnamefont{R.~A.~M.} \bibnamefont{Klomp}},
  \bibinfo{author}{\bibfnamefont{C.~P.~F.} \bibnamefont{Terheggen}},
  \bibnamefont{and} \bibinfo{author}{\bibfnamefont{J.~J.}
  \bibnamefont{de~Swart}}, \bibinfo{journal}{Phys. Rev.}
  \textbf{\bibinfo{volume}{C49}}, \bibinfo{pages}{2950} (\bibinfo{year}{1994}),
  \eprint{nucl-th/9406039}.

\bibitem[{\citenamefont{Holt et~al.}(2004)\citenamefont{Holt, Kuo, Brown, and
  Bogner}}]{Holt:2003rj}
\bibinfo{author}{\bibfnamefont{J.~D.} \bibnamefont{Holt}},
  \bibinfo{author}{\bibfnamefont{T.~T.~S.} \bibnamefont{Kuo}},
  \bibinfo{author}{\bibfnamefont{G.~E.} \bibnamefont{Brown}}, \bibnamefont{and}
  \bibinfo{author}{\bibfnamefont{S.~K.} \bibnamefont{Bogner}},
  \bibinfo{journal}{Nucl. Phys.} \textbf{\bibinfo{volume}{A733}},
  \bibinfo{pages}{153} (\bibinfo{year}{2004}), \eprint{nucl-th/0308036}.

\bibitem[{\citenamefont{{Erkelenz} et~al.}(1971)\citenamefont{{Erkelenz},
  {Alzetta}, and {Holinde}}}]{1971NuPhA.176..413E}
\bibinfo{author}{\bibfnamefont{K.}~\bibnamefont{{Erkelenz}}},
  \bibinfo{author}{\bibfnamefont{R.}~\bibnamefont{{Alzetta}}},
  \bibnamefont{and}
  \bibinfo{author}{\bibfnamefont{K.}~\bibnamefont{{Holinde}}},
  \bibinfo{journal}{Nuclear Physics A} \textbf{\bibinfo{volume}{176}},
  \bibinfo{pages}{413} (\bibinfo{year}{1971}).

\bibitem[{\citenamefont{Epelbaum et~al.}(2000)\citenamefont{Epelbaum, Gloeckle,
  and Meissner}}]{Epelbaum:1999dj}
\bibinfo{author}{\bibfnamefont{E.}~\bibnamefont{Epelbaum}},
  \bibinfo{author}{\bibfnamefont{W.}~\bibnamefont{Gloeckle}}, \bibnamefont{and}
  \bibinfo{author}{\bibfnamefont{U.-G.} \bibnamefont{Meissner}},
  \bibinfo{journal}{Nucl. Phys.} \textbf{\bibinfo{volume}{A671}},
  \bibinfo{pages}{295} (\bibinfo{year}{2000}), \eprint{nucl-th/9910064}.

\bibitem[{\citenamefont{Wiringa et~al.}(1995)\citenamefont{Wiringa, Stoks, and
  Schiavilla}}]{Wiringa:1994wb}
\bibinfo{author}{\bibfnamefont{R.~B.} \bibnamefont{Wiringa}},
  \bibinfo{author}{\bibfnamefont{V.~G.~J.} \bibnamefont{Stoks}},
  \bibnamefont{and}
  \bibinfo{author}{\bibfnamefont{R.}~\bibnamefont{Schiavilla}},
  \bibinfo{journal}{Phys. Rev.} \textbf{\bibinfo{volume}{C51}},
  \bibinfo{pages}{38} (\bibinfo{year}{1995}), \eprint{nucl-th/9408016}.

\bibitem[{\citenamefont{Entem et~al.}(2008)\citenamefont{Entem, Ruiz~Arriola,
  Pavon~Valderrama, and Machleidt}}]{Entem:2007jg}
\bibinfo{author}{\bibfnamefont{D.~R.} \bibnamefont{Entem}},
  \bibinfo{author}{\bibfnamefont{E.}~\bibnamefont{Ruiz~Arriola}},
  \bibinfo{author}{\bibfnamefont{M.}~\bibnamefont{Pavon~Valderrama}},
  \bibnamefont{and}
  \bibinfo{author}{\bibfnamefont{R.}~\bibnamefont{Machleidt}},
  \bibinfo{journal}{Phys. Rev.} \textbf{\bibinfo{volume}{C77}},
  \bibinfo{pages}{044006} (\bibinfo{year}{2008}), \eprint{0709.2770}.

\bibitem[{\citenamefont{Mehen et~al.}(1999)\citenamefont{Mehen, Stewart, and
  Wise}}]{Mehen:1999qs}
\bibinfo{author}{\bibfnamefont{T.}~\bibnamefont{Mehen}},
  \bibinfo{author}{\bibfnamefont{I.~W.} \bibnamefont{Stewart}},
  \bibnamefont{and} \bibinfo{author}{\bibfnamefont{M.~B.} \bibnamefont{Wise}},
  \bibinfo{journal}{Phys. Rev. Lett.} \textbf{\bibinfo{volume}{83}},
  \bibinfo{pages}{931} (\bibinfo{year}{1999}), \eprint{hep-ph/9902370}.

\bibitem[{\citenamefont{Entem and Machleidt}(2003)}]{Entem:2003ft}
\bibinfo{author}{\bibfnamefont{D.~R.} \bibnamefont{Entem}} \bibnamefont{and}
  \bibinfo{author}{\bibfnamefont{R.}~\bibnamefont{Machleidt}},
  \bibinfo{journal}{Phys. Rev.} \textbf{\bibinfo{volume}{C68}},
  \bibinfo{pages}{041001} (\bibinfo{year}{2003}), \eprint{nucl-th/0304018}.

\bibitem[{\citenamefont{Kaplan and Manohar}(1997)}]{Kaplan:1996rk}
\bibinfo{author}{\bibfnamefont{D.~B.} \bibnamefont{Kaplan}} \bibnamefont{and}
  \bibinfo{author}{\bibfnamefont{A.~V.} \bibnamefont{Manohar}},
  \bibinfo{journal}{Phys. Rev.} \textbf{\bibinfo{volume}{C56}},
  \bibinfo{pages}{76} (\bibinfo{year}{1997}), \eprint{nucl-th/9612021}.

\bibitem[{\citenamefont{Amghar and Desplanques}(1995)}]{Amghar:1995av}
\bibinfo{author}{\bibfnamefont{A.}~\bibnamefont{Amghar}} \bibnamefont{and}
  \bibinfo{author}{\bibfnamefont{B.}~\bibnamefont{Desplanques}},
  \bibinfo{journal}{Nucl. Phys.} \textbf{\bibinfo{volume}{A585}},
  \bibinfo{pages}{657} (\bibinfo{year}{1995}).

\bibitem[{\citenamefont{Furnstahl et~al.}(2001)\citenamefont{Furnstahl, Hammer,
  and Tirfessa}}]{Furnstahl:2000we}
\bibinfo{author}{\bibfnamefont{R.~J.} \bibnamefont{Furnstahl}},
  \bibinfo{author}{\bibfnamefont{H.~W.} \bibnamefont{Hammer}},
  \bibnamefont{and} \bibinfo{author}{\bibfnamefont{N.}~\bibnamefont{Tirfessa}},
  \bibinfo{journal}{Nucl. Phys.} \textbf{\bibinfo{volume}{A689}},
  \bibinfo{pages}{846} (\bibinfo{year}{2001}), \eprint{nucl-th/0010078}.

\end{thebibliography}

\end{document}